\documentclass[twocolumn,aps,pra,showpacs,groupedaddress]{revtex4}
\usepackage{epsfig,amssymb,amsmath,mathrsfs}
\usepackage[active]{srcltx}
\usepackage[hypertex,linkcolor=red]{hyperref}
\usepackage[matrix,frame,arrow]{xypic}
\usepackage{color}

\usepackage[matrix,frame,arrow]{xy}
\usepackage{amsmath}
\newcommand{\bra}[1]{\left\langle{#1}\right\vert}
\newcommand{\ket}[1]{\left\vert{#1}\right\rangle}
\newcommand{\qw}[1][-1]{\ar @{-} [0,#1]}
\newcommand{\qwx}[1][-1]{\ar @{-} [#1,0]}
\newcommand{\cw}[1][-1]{\ar @{=} [0,#1]}

\newcommand{\measureD}[1]{*{\xy*+=+<.5em>{\vphantom{\rule{0em}{.1em}#1}}*\cir{r_l};p\save*!R{#1} \restore\save+UC;+UC-<.5em,0em>*!R{\hphantom{#1}}+L **\dir{-} \restore\save+DC;+DC-<.5em,0em>*!R{\hphantom{#1}}+L **\dir{-} \restore\POS+UC-<.5em,0em>*!R{\hphantom{#1}}+L;+DC-<.5em,0em>*!R{\hphantom{#1}}+L **\dir{-} \endxy} \qw}

\newcommand{\multimeasureD}[2]{*+<1em,.9em>{\hphantom{#2}}\save[0,0].[#1,0];p\save !C *{#2},p+LU+<0em,0em>;+RU+<-.8em,0em> **\dir{-}\restore\save +LD;+LU **\dir{-}\restore\save +LD;+RD-<.8em,0em> **\dir{-} \restore\save +RD+<0em,.8em>;+RU-<0em,.8em> **\dir{-} \restore \POS !UR*!UR{\cir<.9em>{r_d}};!DR*!DR{\cir<.9em>{d_l}}\restore \qw}
\newcommand{\control}{*!<0em,.025em>-=-{\bullet}}
\newcommand{\controlo}{*-<.21em,.21em>{\xy *=<.59em>!<0em,-.02em>[o][F]{}\POS!C\endxy}}

\newcommand{\qswap}{*=<0em>{\times} \qw}
\newcommand{\multigate}[2]{*+<1em,.9em>{\hphantom{#2}} \qw \POS[0,0].[#1,0];p !C *{#2},p \save+LU;+RU **\dir{-}\restore\save+RU;+RD **\dir{-}\restore\save+RD;+LD **\dir{-}\restore\save+LD;+LU **\dir{-}\restore}
\newcommand{\ghost}[1]{*+<1em,.9em>{\hphantom{#1}} \qw}

\newcommand{\ustick}[1]{*!D!<0em,-.5em>=<0em>{#1}}

\newcommand{\Qcircuit}[1][0em]{\xymatrix @*[o] @*=<#1>}

\newcommand{\pureghost}[1]{*+<1em,.9em>{\hphantom{#1}}}
\newcommand{\multiprepareC}[2]{*+<1em,.9em>{\hphantom{#2}}\save[0,0].[#1,0];p\save !C
  *{#2},p+RU+<0em,0em>;+LU+<+.8em,0em> **\dir{-}\restore\save +RD;+RU **\dir{-}\restore\save
  +RD;+LD+<.8em,0em> **\dir{-} \restore\save +LD+<0em,.8em>;+LU-<0em,.8em> **\dir{-} \restore \POS
  !UL*!UL{\cir<.9em>{u_r}};!DL*!DL{\cir<.9em>{l_u}}\restore}

\newtheorem{lemma}{Lemma}

\newtheorem{remark}{Remark}
\def\d{\operatorname d}
\def\vec#1{\boldsymbol{#1}}
\def\qed{$\blacksquare$}
 
\def\>{\rangle}
\def\<{\langle}\def\kk{\>\! \>}          
\def\Ket#1{|#1\>}

  \def\Tr{{\rm Tr}}
 
\def\ketbra#1#2{|#1\>\<#2|}
\def\group#1{\mathbb{#1}}
\def\hilb#1{\mathscr{#1}}
\def\defset#1{\mathsf{#1}}
\def\Rel#1{C^{#1}_d}

\newcommand{\multipuregate}[2]{*+<1em,.9em>{\hphantom{#2}}  \POS[0,0].[#1,0];p !C *{#2},p \save+LU;+RU **\dir{-}\restore\save+RU;+RD **\dir{-}\restore\save+RD;+LD **\dir{-}\restore\save+LD;+LU **\dir{-}\restore}

\begin{document}

\title{Quantum learning algorithms for quantum measurements}
\author{Alessandro Bisio}\email{alessandro.bisio@unipv.it}
\affiliation{{\em QUIT Group}, Dipartimento di Fisica  ``A. Volta'' and INFN, via Bassi 6, 27100 Pavia, Italy}
\homepage{http://www.qubit.it}
\author{Giacomo Mauro D'Ariano}\email{dariano@unipv.it}
\affiliation{{\em QUIT Group}, Dipartimento di Fisica  ``A. Volta'' and INFN, via Bassi 6, 27100 Pavia, Italy}
\homepage{http://www.qubit.it}
\author{Paolo Perinotti}\email{paolo.perinotti@unipv.it}
\affiliation{{\em QUIT Group}, Dipartimento di Fisica  ``A. Volta'' and INFN, via Bassi 6, 27100 Pavia, Italy}
\homepage{http://www.qubit.it}
\author{Michal Sedl\'ak}\email{michal.sedlak@unipv.it}
\affiliation{{\em QUIT Group}, Dipartimento di Fisica  ``A. Volta'', via Bassi 6, 27100 Pavia, Italy}
\affiliation{Institute of Physics, Slovak Academy of Sciences, D\'ubravsk\'a cesta 9, 845 11 Bratislava, Slovakia}
\homepage{http://www.qubit.it}
\date{\today}
\begin{abstract}
  We study quantum learning algorithms for quantum measurements. The
  optimal learning algorithm is derived for arbitrary von Neumann
  measurements in the case of training with one or two examples. The analysis of the
  case of three examples reveals that, differently from the learning
  of unitary gates, the optimal algorithm for learning of quantum
  measurements cannot be parallelized, and requires quantum memories
  for the storage of information.
\end{abstract}
\pacs{03.67.-a, 03.67.Ac, 03.65.Ta}\maketitle

\section{Introduction}


The rapid development of an information
technology in the last decades made the optimization of information processing tasks an important field of computer science.
For example one needs to optimize database search, as well as tasks that emerged due to internet e.g. algorithms for anti-spam filters and internet search engines. The last two tasks are instances of the so called machine learning \cite{vapnik}, which can be defined as follows. Suppose we have a black box evaluating an unknown function $f$ and we have access to $N$ uses of it. However, after we lose the access to the black box we need to evaluate $f$ on an input that was not previously available. Naturally any machine learning has two phases -- training and retrieving. The knowledge on $f$ acquired in the training phase of the strategy is encoded into a bit string that is later used as a program governing the retrieval phase. Obviously, if $N$ is greater or equal to the number of possible inputs of $f$ then the training part of the strategy can acquire complete knowledge of $f$.
The same task, termed quantum learning, can be generalized to quantum theory. In this case the black box performs an unknown quantum transformation $\mathcal{T}$. 
The result of the training phase is a quantum state $\psi_{\mathcal{T}}$. This state has to be kept in the quantum memory until the retrieving phase, where it enters together with the unknown state $\rho$ into the retrieving channel that mimics the action of $\mathcal{T}$ on $\rho$.
We can immediately observe substantial difference to  machine learning. Even for finite dimensional quantum systems there does not exist a finite $N$ for which the quantum learning works perfectly. Indeed, even if the training part of the strategy would encode full information about $\mathcal{T}$ into the finite dimensional state $\psi_{\mathcal{T}}$, the no programming theorem of Nielsen \cite{nielsen} prevents us to retrieve the transformation perfectly.



A closely related problem to quantum learning was studied as a quantum version of
pattern recognition algorithms \cite{sasa,sasacarl}. For the case of quantum learning of channels,
the first analysis was published in Ref. \cite{molmer}, where very simple processing techniques were studied
for learning of particular gates like the
Grover oracle \cite{Grover} or the discrete Fourier transform. Learning of unitary black boxes was analyzed in Ref.
\cite{learnunit}.  Surprisingly, it turns out that the task of quantum
learning of unitaries can be fully parallelized, which means that the
optimal training phase is achieved by applying the $N$ uses of the black box on the fixed entangled state.
Another surprising feature of the aforementioned training phase is that it is an optimal estimation procedure and hence the quantum memory can be replaced by a classical storage of the estimated unitary black box. The simulation then consists in the conditional application of
the gate corresponding to the estimated parameters.

In the present paper we will consider the case
in which the black box to be learnt is a device performing a
Von Neumann measurement, namely a projective non-degenerate
Positive Operator Valued Measure (POVM) $\mathbf{E} := \{ E_i \}$.
We will 
show that for measuring black boxes the 
surprising features of optimal learning of unitary black boxes disappear. In particular, we
will show that the optimal algorithm cannot be parallelized, leading
to a training phase that lasts an increasing time versus the number of
examples.
Moreover, the optimal training does not consist of optimal estimation, thus
requiring a coherent quantum memory for the storage of the learnt
measurement.

The paper is organized as follows. In Sec \ref{prelim} we review some
notation and preliminary concepts used in the analysis. In Sec.
\ref{mathform} we expose the mathematical formulation of the general
problem of optimal learning  in mathematical
terms. In Sec.  \ref{sec:obssymm-repl-netw} the problem is simplified
exploiting all the symmetries that can be useful. The problem is then
solved in Sec.  \ref{learn} for the cases $N = 1$,
$N = 2$ and $N = 3$.
Finally, the paper is closed by concluding remarks in Sec. \ref{conc}.

\section{Preliminary concepts}\label{prelim}

In this section we review some notions of the theory of \emph{quantum
  networks} \cite{comblong,architecture,memoryeff}.  The main feature of
this approach is the representation of quantum networks in terms of
suitably normalized positive operators.

The nodes of a quantum network $\mathcal{R}$ are elementary boxes
linked by wires. Elementary boxes represent state preparations,
channels, quantum operations, or effects. The most general pictorial
representation of a quantum network is a directed acyclic graph, where
the vertices represent elementary boxes and the arrows represent
the quantum systems traveling within the network in the direction
induced by the input-output relation.

By stretching the connections in the graph we can give the quantum
network the shape of a comb, i.e. any quantum network $\mathcal{R}$ is
equivalent to a sequence of $N$ quantum operations $\{T_i\}_{i=1}^N$
with some unconnected input and output subsystems, as follows
\begin{equation}
  \Qcircuit @C=1em @R=1em {
    \ustick{0}&\multigate{1}{T_1}&\ustick{1}\qw&&\ustick{2}\qw&\multigate{1}{T_2}&\ustick{3}\qw&\ghost{\dots}&\ustick{2N-2}\qw&\multigate{1}{T_N}&\ustick{2N-1}\qw\\
    &\pureghost{T_1}&\qw&\qw&\qw&\ghost{T_2}&\qw&\cdots&&\ghost{T_N}&}
  \label{scheme}
\end{equation}
If all the $N$ quantum operations are trace preserving (i.e.  they are
quantum channels) $\mathcal{R}$ is a {\em deterministic} quantum
network, otherwise $\mathcal{R}$ is a {\em probabilistic} quantum
network.  The ordering of the teeth is induced by the causal order
defined by the flow of quantum information inside the quantum network.
Referring to the scheme in Eq.~\eqref{scheme} we label each wire with
an integer number $j$: accordingly, the Hilbert space of the system
represented by wire $j$ is denoted as $\hilb{H}_j$.

Since a quantum network $\mathcal{R}$ is a concatenation of quantum
operations it can be considered as a quantum operation itself
$\mathcal{R}:\mathcal{L}(\hilb{H}_{even}) \to
\mathcal{L}(\hilb{H}_{odd})$ where we defined $\hilb{H}_{even}=
\bigotimes_{i=0}^N \hilb{H}_{2i}$ and $\hilb{H}_{odd}=
\bigotimes_{i=0}^N \hilb{H}_{2i+1}$.  That being so, it is possible to
define the Choi-Jamio\l kowsky operator of a quantum network as
\begin{align}
& R :=  \mathcal{R} \otimes \mathcal{I} (\ketbra{\omega}{\omega}) \\
&  R \in \mathcal{L}(\hilb{H}_{even}\otimes \hilb{H}_{odd} ), \qquad R \leq 0 \nonumber
\end{align}
where $\mathcal{I}$ is the identity map and
$\ket{\omega} \in \hilb{H}_{even}^{\otimes 2}$, $\ket{\omega} = \sum_{n}\ket{n}\ket{n}$
($\{ \ket{n}  \} $ is an orthonormal basis of $\hilb{H}_{even}$).
The Choi-Jamio\l kowsky operator of a quantum network is called
\emph{quantum comb} of the network.
If $\mathcal{R}$
is a deterministic quantum network
it is possible to prove that its Choi-Jamio\l kowsky operator $R$
must satisfy the recursive normalization constraint
\begin{align}\label{recnorm}
\Tr_{2k-1} [ R^{(k)}] = I_{2k-2} \otimes R^{(k-1)} \qquad k=1, \dots, N~
\end{align}
where $R^{(N)}=R$, $R^{(0)} =1$, $R^{(k)}
\in \mathcal{L}(\hilb{H}_{{odd}_k} \otimes \hilb{H}_{{even}_k})$
with
 $\hilb{H}_{{even}_k} = \bigotimes_{j=0}^{k-1} \hilb{H}_{2j}$ and
$\hilb{H}_{{odd}_k} = \bigotimes_{j=0}^{k-1} \hilb{H}_{2j+1}$,
is the comb of the reduced circuit $\mathcal R^{(k)}$ obtained by
discarding the last $N-k$ teeth.
It is relevant to stress that each positive operator that
satisfies Eq. (\ref{recnorm}) corresponds to a valid deterministic quantum network.
This gives us a correspondence between the set of positive operators
satisfying Eq. (\ref{recnorm}) and the set of deterministic quantum networks.

On the other hand, the Choi-Jamio\l kowsky operator of
 a probabilistic quantum network $\mathcal{R}$,
must satisfy
\begin{align}\label{eq:recnormprob}
  0 \leq R \leq S
\end{align}
where $S$ is a Choi-Jamio\l kowsky operator of a deterministic
quantum network. An important theorem proves \cite{comblong} that any
positive operator, upon suitable rescaling, represents a probabilistic
quantum network.

Two quantum networks $\mathcal{R}_1$ and $\mathcal{R}_2$ can be
connected by linking input wires of one network with output wires of
the other network thus forming the network $\mathcal{R}_1 *
\mathcal{R}_2$. The Choi-Jamio\l kowsky operator of the composite
network $\mathcal{R}_1 * \mathcal{R}_2$ is the \emph{link product} of
the operators $R_1$ and $R_2$ which is defined as follows:
\begin{align}
  R_1*R_2 = \Tr_{\mathcal{K}}[R_1 R_2^{\theta_{\mathcal{K}}}]
\end{align}
where $\theta_{\mathcal{K}}$ denotes the partial transposition (with
respect to a fixed orthonormal basis) over the Hilbert space
$\mathcal{K}$ of the connected wires and $\Tr_{\mathcal{K}}$ denotes
the partial trace over $\mathcal{K}$.

\subsection{Generalized Instrument}

The aim of this paper is to study quantum networks that replicate
quantum measurements.  A generalized quantum instrument is set of
probabilistic quantum networks $\boldsymbol{\mathcal{R}} := \{
\mathcal{R}_i \}$
 such that
the set $\mathbf R = \{ R_i\}$ of the Choi-Jamio\l kowsky operators of its components satisfies the
following condition:
\begin{align}
  &\sum_i R_i := R_{\Omega}
\end{align}
where $R_{\Omega}$ corresponds to a deterministic quantum network.  Every
probabilistic quantum network belongs to some generalized quantum
instrument, and viceversa every generalized quantum instrument
represents some set of probabilistic quantum networks.

\section{the optimization problem}\label{mathform}

The learning scenario can be formulated as a quantum network
that accepts $N$ measurements
into the open slots and works as a POVM
on the remaining system.
 Here is a diagram representing the $N=2$ scenario,
\begin{align}\label{eq:obs21learn}
  \begin{aligned}
    \Qcircuit @C=0.7em @R=1em { &&&&&&& \ustick{4}& \ghost{\;\;\;}\\
      \multiprepareC{1}{\;\;\;} &\ustick{0} \qw & \measureD{\mathbf{E}} &
      \ustick{1} \cw & \pureghost{\;\;\;} \cw & \ustick{2} \qw &
      \measureD{\mathbf{E}} & \ustick{3} \cw &
      \pureghost{\;\;\;} \cw \\
      \pureghost{\;\;\;}& \qw& \qw&\qw & \multigate{-1}{\;\;\;}&\qw &
      \qw&\qw &
      \multigate{-2}{\;\;\;} \\
    }
  \end{aligned}
\simeq\begin{aligned} \Qcircuit @C=0.5em @R=0.5em {
      \ustick{}&\measureD{\mathbf{E}}\\
      }
  \end{aligned}
\nonumber
\end{align}
where the double wires carry the classical outcomes of the
measurements.

Since we consider the case where the unknown measurement is a
projective non degenerate POVM $\mathbf {E}:=\{E_1,\dots,E_d\}$, we can write
its element $E_i$ in the following form
\begin{align}
  E_i = \ketbra{\phi_i}{\phi_i}
\end{align}
where $\{ \ket{\phi_i} \}_{i=1}^d$ is an orthonormal basis of the Hilbert space
$\mathcal{H}$.  All the POVM's of this kind can be generated by
rotating a reference POVM $\mathbf {E}:=\{|i\>\<i|\}_{i=1}^d$ by 
elements of the group of unitary transformations $\group{SU}(d)$ as
follows
\begin{align}
  \mathbf E^{(U)} := U\mathbf E U^\dagger \qquad U \in \group{SU}(d),
\end{align}
where $\{\ket{i}\}$ is a fixed orthonormal basis and $U\mathbf E U^\dagger$ denotes the POVM with elements
$E^{(U)}_i:=UE_iU^\dagger$. Notice the slight abuse in the definition
of $\mathbf{E}^{(U)}$, due to the fact that there exists a stability
subgroup $\mathsf{S}\subseteq\mathbf{SU}(d)$ such that for
$V\in\mathsf S$ one has $V|i\>=|i\>$ for all $i$. The POVM $\mathbf
{E}^{(U)}$ is then labeled by the equivalence class $[U]$ defined by
the relation
\begin{equation}
  U\sim U'\quad\Leftrightarrow\quad U=U'V,\quad V\in \mathsf S,
\end{equation}
rather than by $U$.


It is formally convenient to encode the classical outcome $i$ of the POVM 
into a quantum system by preparing the state $\ket{i}$ from a
fixed orthonormal basis, which is the same for each POVM \cite{directsum}.
Within this framework the measurement device is actually
described by the following measure-and-prepare quantum channel
$\mathcal{E}^{(U)}:\mathcal{L}(\hilb{H})\to \mathcal{L}(\hilb{H})$
\begin{align}
  \mathcal{E}^{(U)}(\rho) = \sum_{i=1}^d\Tr[E_i^{(U)}\rho]\ketbra{i}{i},
\end{align}
which measures the POVM $\mathbf {E}^{(U)}$ on {\bf the} input state and
in the case of outcome $i$ prepares the state $\ket{i}$ from a fixed orthonormal basis on the
output of the channel.
The Choi-Jamio\l kowski representation of the channel $\mathcal{E}^{(U)}$
is the following
\begin{align}\label{eq:obsdepolachan}
  {E}^{(U)} = \sum_{i=1}^d \ketbra{i}{i} \otimes {E_i^{(U)}}^T = \sum_{i=1}^d \ketbra{i}{i}\otimes U^* \ketbra{i}{i} U^T,
\end{align}
where $X^T$ denotes the transpose of $X$ in the basis $\{\ket
i\}_{i=1}^d$. The $N$ uses of the measurement device are then
represented by the tensor product $E^{(U)}_{2N-1 \; 2N-2} \otimes
\cdots \otimes E^{(U)}_{10} $ where the input and the output space of
the $k$-th use of the measurement device are denoted by $2k-2$ and
$2k-1$, respectively.  We introduce the following notation
\begin{align}
  \hilb{H}_{\defset{in}} := \bigotimes_{k=1}^N \hilb{H}_{2k-2}, \qquad
  \hilb{H}_{\defset{cl}} := \bigotimes_{k=1}^N \hilb{H}_{2k-1} .
\end{align}
Since we want the learning network $\boldsymbol{\mathcal{R}} $ to
behave as the 
POVM $\mathbf{E}^{(U)}$ upon insertion of the $N$ uses of
$\mathcal{E}^{(U)}$, we have that
$\mathbf R$ is a
generalized instrument where the element $R_i$ describes
the behaviour of the network when the output
of the replicated measurement
is $i$. The replicated POVM
is then equal to
\begin{align}\label{eq:obsfinalpovm}
  &\mathbf{G}^{(U)} = {[ \mathbf{R}*( {E}_{2N-1 \; 2N-2}^{(U)} * \cdots *{E}_{10}^{(U)})]}^T \\
  \nonumber &R_i = \mathcal{L}( \hilb{H}_{\defset{out}} \otimes
  \hilb{H}_{\defset{cl}} \otimes\hilb{H}_{\defset{in}}), \quad
  \hilb{H}_{\defset{out}} = \hilb{H}_{2N}
\end{align}
where $\hilb{H}_{2N}$ denotes the input space of the replicated
measurement.
In this notation the normalization of the generalized instrument $\vec R$ becomes
 \begin{align}
   &\Tr_{2k-2}[R^{(k)}]=I_{2k-3}\otimes R^{(k-1)},\quad k=1,\ldots,N \nonumber\\
   &R_\Omega=I_{2N,2N-1}\otimes R^{(N)},\quad R^{(0)}=1.
   \label{eq:recnorm2}
 \end{align}

Our task is to find the learning network 
$\boldsymbol{\mathcal{R}}$ such that $\mathbf{G}^{(U)}$ is as close as
possible to $\mathbf {E}^{(U)}$.  In order to quantify the performances
of the replicating network, we introduce the following quantity that
measures the closeness between two POVM's $\mathbf P$ and $\mathbf Q$
\begin{equation}
   \mathscr{D}(\mathbf P,\mathbf Q):= \int d\psi \sum_{i=1}^d |\bra{\psi}P_i-Q_i\ket{\psi}|^2
   \label{eq:distance}
\end{equation}
The interpretation of $\mathscr{D}(\mathbf P,\mathbf Q)$ as a measure of "distance"
between $\mathbf{P}$ and $\mathbf{Q}$ is provided by the following
Lemma.

\begin{lemma}[Distance criterion for two POVMs]
  Let $\Sigma := \{ 1,\dots,d \}$ be a finite set of events and
$ \mathbf{P} \subseteq \mathcal{L}(\hilb{H})$ and $\mathbf{Q}
  \subseteq\mathcal{L}(\hilb{H}) $ be two POVM's.
  Consider now the quantity $\mathscr{D}(\mathbf P,\mathbf Q)$ from equation (\ref{eq:distance})
  Then the following properties hold: \\
  i) $\mathscr{D}(\mathbf P,\mathbf Q)\geq 0$, \\
  ii) $\mathscr{D}(\mathbf P,\mathbf Q)=0 \Leftrightarrow P_i = Q_i\ \;  \forall i$, \\
  iii) $\mathscr{D}(\mathbf P,\mathbf Q)$ is convex with respect to POVMs. \\
  iv) $\mathscr{D}(U \mathbf P U^\dagger,U \mathbf Q U^\dagger)=\mathscr{D}(\mathbf P,\mathbf Q)$ for any unitary operator $U$.
\end{lemma}

\begin{Proof}
  The non negativity of function $f(x)=x^2$ guarantees the same
  property also for $\mathscr{D}$, which is a sum and an integral of
  the squares. 
  For $P_i = Q_i \;\forall i$ it is obvious that $\mathscr{D}(\mathbf
  P,\mathbf Q)=0$. To prove the converse, it suffices to realize that
  $\mathscr{D}(\mathbf P,\mathbf Q)=0$ implies
  $\bra{\psi}P_i-Q_i\ket{\psi}=0 \;\forall \psi$, which by
  polarization identity requires $P_i = Q_i \;\forall i$.  In order to
  prove convexity, we need to show that
  \begin{eqnarray}
  & &\mathscr{D}(\mathbf P,\lambda \mathbf Q + (1-\lambda)\mathbf Q') \leq    \\
  & &\quad\quad\leq \lambda \mathscr{D}(\mathbf P,\mathbf Q) + (1-\lambda)\mathscr{D}(\mathbf P,\mathbf{Q}')  \nonumber
  \end{eqnarray}
  holds for any POVM $\mathbf{Q}'$ and $0\leq \lambda \leq 1$. If we denote $a_i=\bra{\psi}P_i-Q_i\ket{\psi}$, $b_i=\bra{\psi}P_i-Q'_i\ket{\psi}$ and utilize convexity of $f(x)=x^2$, i.e.
  \begin{eqnarray}
    (\lambda a_i +(1-\lambda)b_i)^2\leq \lambda a^2_i +(1-\lambda)b^2_i \nonumber
  \end{eqnarray}
  then the claim follows directly from the definition in Eq. (\ref{eq:distance}). Similarly, property iv) is obvious from the definition in Eq. (\ref{eq:distance}).
  \qed
\end{Proof}

Assuming that the unknown POVM $\mathbf E^{(U)}$ is randomly drawn
according to the Haar distribution, we choose the quantity:
\begin{align}\label{eq:obsfigmer}
  D := \int \d U \mathscr{D}(\mathbf E^{(U)},\mathbf G^{(U)})
\end{align}
as a figure of merit for the learning network. The quantity $D$
clearly depends on the network $\boldsymbol{\mathcal R}$, and will be
denoted by $D[\boldsymbol{\mathcal R}]$. Our task is to find the
optimal generalized instrument $\vec R$, that
minimizes $D[\boldsymbol{\mathcal R}]$.

\section{Symmetries of the learning network}\label{sec:obssymm-repl-netw}

In this section we utilize 
the symmetries of the figure of merit
(\ref{eq:obsfigmer}) to simplify the optimization problem. The first
simplification relies on the fact that some wires of the network carry
only classical information, representing the outcome of the
measurement.

\begin{lemma}[Restriction to diagonal network]\label{lem:diagnet}
  The optimal generalized instrument $\vec {R}$, $\sum_{i}
  R_{i} = R_{\Omega}$ minimizing Eq.
  (\ref{eq:obsfigmer}) can be chosen to satisfy:
\begin{align}\label{eq:obsdiagcomb}
  R_{i} = \sum_{\vec j}  R'_{i,\vec{j} } \otimes \ketbra{\vec{j}}{\vec{j}},
\end{align}
where $\vec{j} = (j_1, \dots, j_N)$, $\ket{\vec{j}}:=
\ket{j_1}_1\otimes \cdots \otimes \ket{j_N}_{2N-1} \in
\hilb{H}_{\defset{cl}}$, $0\le R'_{i,\vec{j}} \in \mathcal{L}
(\hilb{H}_{\defset{out}} \otimes \hilb{H}_{\defset{in}})$, and
$\sum_{\vec j}$ is a shorthand for $\sum_{j_1,\dots,j_N=1}^d$.
\end{lemma}
\begin{Proof}
Let $\vec S$ be a generalized instrument corresponding to a quantum network $\vec {\mathcal{S}}$.
  Let us define set of operators $\vec {R}$ as 
  \begin{align}
    {R}_{i} := \sum_{\vec j} R'_{i,\vec j} \otimes \ketbra{\vec{j}}{\vec{j}},
  \end{align}
  with $R'_{i,\vec j}:=\bra{\vec{j}} S_{i}\ket{\vec{j}}$. We can
  easily prove that $\vec{R}$ is a generalized instrument.
  Indeed, reminding Eq.  (\ref{eq:obsdepolachan}), we have
  \begin{align}
    \sum_{i} {R}_{i} =& \sum_{i} \sum_{\vec j}
    \bra{\vec{j}} S_{i}\ket{\vec{j}} \otimes
    \ketbra{\vec{j}}{\vec{j}} =\nonumber\\
    &\sum_{\vec j} \bra{\vec{j}} S_{\Omega}\ket{\vec{j}} \otimes \ketbra{\vec{j}}{\vec{j}}=\nonumber\\
    &S_\Omega * E^{(I)} * \cdots * E^{(I)},\label{decchoi}
  \end{align}
  where the link is performed only on the space $\hilb{H}_{\defset{cl}}$.
  The operator in Eq.~\eqref{decchoi} is the Choi-Jamio\l kowski operator of a
  deterministic quantum network satisfying the same normalization
  conditions as $S_\Omega$.  Finally we show that $\vec S$ 
  and  $\vec {R}$ 
  produce the same 
  replicated POVM $\mathbf G^{(U)}$
  when linked with the $N$ uses of $E^{(U)}$, as follows
  \begin{align} \label{eq:povmform}
    (G^{(U)}_{i})^T &= {S_{i}* {E}_{2N-1 \; 2N-2}^{(U)} * \cdots *{E}_{10}^{(U)}} = \nonumber \\
    & \sum_{\vec{j}} ( \bra{\vec{j}}_{\defset{cl}}\bra{\vec{j}}_{\defset{in}}
      {U^\dagger}^{\otimes N})
      S_{i} ( \ket{\vec{j}}_{\defset{cl}}U^{\otimes N}  \ket{\vec{j}}_{\defset{in}}) = \nonumber \\
    &  \sum_{\vec{j}} ( \bra{\vec{j}}_{\defset{in}}
      {U^\dagger}^{\otimes N})
      R'_{i,\vec j} ( U^{\otimes N}  \ket{\vec{j}}_{\defset{in}}) =  \nonumber \\
    & R_{ i}* {E}_{2N-1 \;
        2N-2}^{(U)} * \cdots *{E}_{10}^{(U)}.
  \end{align}
\qed
\end{Proof}

It is clear from Eq.~\eqref{eq:povmform} that also for non diagonal
networks $\boldsymbol{\mathcal R}$, the only relevant terms of the
generalized instrument both for its normalization and for the figure
of merit $D[\boldsymbol{\mathcal R}]$ are
\begin{eqnarray}
R'_{i,\vec j}:=\<\vec j|_{\defset{cl}}R_i|\vec j\>_{\defset{cl}}.
\label{eq:diagpart}
\end{eqnarray}
In the following we will 
use the above notation also for general networks. As a next step, we       %
introduce a unitary symmetry of the learning network and we study its
consequences on the form of the replicated POVM. We will show that
restriction to covariant learning networks can be made without loss of
generality.  For this purpose we introduce the following lemma.

\begin{lemma}[Covariant networks]\label{lem:covnet}
  The optimal generalized instrument $\vec {R}$, $\sum_{i}
  R_{i} = R_{\Omega}$ minimizing Eq.  (\ref{eq:obsfigmer}) can be
  chosen to satisfy
  \begin{align}\label{eq:obscommutr}
    [R_{i}, {U^*}_{\defset{out}} \otimes U^{\otimes
      N}_{\defset{in}}\otimes I_{\defset{cl}}]=0.
  \end{align}
  Then the replicated POVM for $\vec{R}$ enjoys the following
  property
  \begin{eqnarray}
    \mathbf G^{(U)} &=& U\; \mathbf G^{(I)}\;U^\dagger. \label{eq:covpovmform2}
  \end{eqnarray}
\end{lemma}

\begin{Proof}
  From an arbitrary learning network $\vec {\mathcal{S}}$ 
  by symmetrization, we can define a covariant learning network $\vec
  {\mathcal{R}}$ as follows
  \begin{equation}
    R_{i}:=\int \d U ({U^*}\otimes{U}^{\otimes N}\otimes I_{\defset{cl}})S_{i}({U^T}\otimes{U^\dagger}^{\otimes N}\otimes I_{\defset{cl}}).
    \label{eq:symr}
  \end{equation}
  It is easy to verify that the set $\vec{R}$ defines a
  generalized instrument. Moreover, By the invariance of the Haar
  measure $\d U$, the elements of $\vec {R}$ obey
  Eq.~\eqref{eq:obscommutr}. First we show that the replicated POVM
  for the symmetrized instrument $\vec{R}$ enjoys the property (\ref{eq:covpovmform2}).
  Indeed, eq. (\ref{eq:povmform}) 
  provides the following formula for the replicated POVM
  \begin{align}
    (G^{(U)}_{i})^T &= \sum_{\vec{j}} \bra{\vec{j}}_{\defset{in}}
    {U^\dagger}^{\otimes N} R'_{i,\vec j}U^{\otimes N}
    \ket{\vec{j}}_{\defset{in}}, \label{eq:reppovmdiagr}
  \end{align}
  and exploiting the expression in Eq. (\ref{eq:symr}) for $R'_{i,\vec
    j}$, we obtain
  \begin{eqnarray}
    \mathbf G^{(U)} &=& \int dW\; W \mathbf Q^{(W^\dagger U)} W^\dagger,  \label{eq:covpovmform1}
  \end{eqnarray}
  where $\mathbf Q^{(U)}$ denotes the replicated POVM for the original
  learning network $\vec {\mathcal{S}}$. Eq.  (\ref{eq:covpovmform2})
  is a direct consequence of Eq.  (\ref{eq:covpovmform1}), which can
  be seen via suitable shift of the invariant Haar integration
  measure. We can now show that $D[\boldsymbol{\mathcal R}]\leq
  D[\boldsymbol{\mathcal S}]$ as follows
  \begin{eqnarray}
    D[\boldsymbol{\mathcal R}]&=&\int \d U \mathscr{D}(\mathbf E^{(U)},\int dW\; W \mathbf Q^{(W^\dagger U)} W^\dagger) \nonumber \\
    &\leq&\int \d W \d U   \mathscr{D}(U\mathbf E U^\dagger,W \mathbf Q^{(W^\dagger U)} W^\dagger)  \nonumber \\
    &\leq&\int \d U  dW \mathscr{D}(W \mathbf E^{(U)} W^\dagger,W \mathbf Q^{(U)} W^\dagger)  \nonumber \\
    &=&D[\boldsymbol{\mathcal S}], \nonumber
  \end{eqnarray}
  where we used properties iii), iv) of $\mathscr{D}$ and shifted the
  Haar invariant integration measure $dU$ to $d(W^\dagger U)$.  \qed
\end{Proof}

Another symmetry we introduce is related to the possibility
of relabeling the outcomes of a POVM. We shall denote by $\sigma$ the
element of $\group{S}_d$, the group of permutations of $d$ elements,
and by $T_\sigma$ the linear operator that permutes the elements of basis
$\{\ket{i}\}$ according to this permutation, in formula
$T_\sigma\ket{i}=\ket{\sigma(i)}$. Let us note that the complex conjugation and transposition are defined with respect to the basis $\{\ket{i}\}$, so $T_\sigma=T^*_\sigma$.

\begin{lemma}[Relabeling symmetry]
  \label{lem:relabsym}
  The optimal covariant generalized instrument $\vec R$,
  $\sum_{i} R_{i} = R_{\Omega}$ minimizing Eq.  (\ref{eq:obsfigmer})
  can be chosen to satisfy Eq.~\eqref{eq:obscommutr} and the following
  condition
  \begin{align}\label{eq:obsperminvproperty}
    R_{i} = (I_{\defset{out}}\otimes I_{\defset{in}}\otimes
    {T^T_\sigma}^{\otimes N})R_{\sigma(i)}(I_{\defset{out}}\otimes I_{\defset{in}}\otimes {T_\sigma}^{\otimes N}),
  \end{align}
  where $\sigma(\vec{j}):=(\sigma(j_1),\ldots,\sigma(j_M))$. Then the
  seed of replicated POVM satisfies
  \begin{align}\label{eq:obspermseedproperty}
    \mathbf G^{(I)}_\sigma = T_\sigma \mathbf G^{(I)}T_\sigma^\dagger
     \qquad \forall \sigma \in \group{S}_d.
  \end{align}
  where $\mathbf X_\sigma$ denotes the ordered set with elements
  $(X_\sigma)_i:=X_{\sigma(i)}$.
\end{lemma}

\begin{Proof}
For a given covariant learning network $\boldsymbol{\mathcal S}$
  satisfying Eq.~\eqref{eq:obscommutr}, let us define
  \begin{align}
    R_{i}:=&\frac{1}{d!}  \sum_{\sigma\in \group{S}_d}
    (T_\sigma\otimes T_\sigma^{\otimes N}\otimes
    T_\sigma^{\otimes N})^T S_{\sigma(i)}(T_\sigma\otimes
    {T_\sigma}^{\otimes N}\otimes {T_\sigma}^{\otimes N}),\nonumber\\
    =& \frac{1}{d!} \sum_{\sigma\in \group{S}_d}(I_{\defset{out}}\otimes I_{\defset{in}}\otimes
    {T^T_\sigma}^{\otimes N}) S_{\sigma(i)}(I_{\defset{out}}\otimes I_{\defset{in}}\otimes
    {T_\sigma}^{\otimes N}),
    \label{eq:obscompatsym}
  \end{align}
  where the last identity follows from the commutation relation
  (\ref{eq:obscommutr}) with $ U = T^T_\sigma$. The generalized
  instrument $\mathbf R$ corresponds to a covariant quantum network
  $\vec{\mathcal{R}}$, because it represents a convex combination of
  well-normalized covariant networks.
  The quantum network $\vec{\mathcal{R}}$ operationally corresponds to a
  random simultaneous relabeling of the outcomes of the inserted and
  replicated measurements by permutation $\sigma$. Let us now prove
  Eq. (\ref{eq:obspermseedproperty}). 

Since generalized instrument $\vec R$ inherits commutation property
(\ref{eq:obscommutr}) from $\vec S$ (see definition
(\ref{eq:obscompatsym})) it is obvious that the introduced permutation
symmetry will not spoil the existing covariance from Eq.
(\ref{eq:covpovmform2}). Thus, it suffice to investigate how the seed
of the replicated POVM changes, when we introduce permutation
symmetry.

  Inserting definition (\ref{eq:obscompatsym}) into Eq.
  (\ref{eq:povmform}) we find
  \begin{eqnarray}
    (G^{(I)}_{i})^T
    &=&\frac{1}{d!} \sum_{\sigma\in \group{S}_d} T^T_{\sigma} \sum_{\vec{j}}   \bra{\sigma(\vec{j})}_{\defset{in}}S'_{\sigma(i),\sigma(\vec{j})}\ket{\sigma(\vec{j})}_{\defset{in}} T^*_{\sigma} \nonumber\\
    &=&\frac{1}{d!} \sum_{\sigma\in \group{S}_d} T^T_{\sigma} \sum_{\vec{j}}  \bra{\vec{j}}_{\defset{in}}
    S'_{\sigma(i),\vec{j}}\ket{\vec{j}}_{\defset{in}} T^*_{\sigma} \nonumber\\
    &=&\frac{1}{d!} \sum_{\sigma\in \group{S}_d}  T^T_{\sigma}   (Q^{(I)}_{\sigma(i)})^T  T^*_{\sigma},
 \label{eq:permutseed}
  \end{eqnarray}
  where we defined $S'_{i,\vec j}:=\<\vec j|S_i|\vec
  j\>$, and
  we denoted by $\mathbf Q^{(U)}$ the POVM replicated by the
  original learning network $\vec{\mathcal{S}}$.  Transposing the last
  equation one can easily
  derive Eq.  (\ref{eq:obspermseedproperty}) by analyzing the
  conjugation with $T^T_\tau$ $\tau\in\group{S}_d$.

As a next step, we show that $D[\boldsymbol{\mathcal R}]\leq
D[\boldsymbol{\mathcal S}]$. Indeed,
\begin{eqnarray}
  D[\boldsymbol{\mathcal R}]&=&\int \d U \mathscr{D}(\mathbf E^{(U)},\frac{1}{d!} \sum_{\sigma\in \group{S}_d}
  UT^\dagger_\sigma \mathbf Q^{(I)}_{\sigma} T_\sigma U^\dagger ) \nonumber \\
  &\leq&\frac{1}{d!} \sum_{\sigma\in \group{S}_d}\int \d U \mathscr{D}(\mathbf E_{\sigma}^{(UT^\dagger_\sigma)},
   \mathbf Q^{(UT^\dagger_\sigma)}_{\sigma} ) \nonumber \\
  &\leq&\frac{1}{d!} \sum_{\sigma\in \group{S}_d}\int \d W \mathscr{D}(\mathbf E_{\sigma}^{(W)},
  \mathbf Q^{(W)}_{\sigma}) \nonumber \\
  &\leq&D[\boldsymbol{\mathcal S}], \nonumber
\end{eqnarray}
where we utilized Eq.  (\ref{eq:covpovmform2}), convexity of    
$\mathscr{D}(\mathbf E^{(U)},\mathbf G^{(U)})$, and the fact that
$\mathscr{D}(\mathbf E_{\sigma}^{(U)},\mathbf
Q_{\sigma}^{(U)})=\mathscr{D}(\mathbf E^{(U)},\mathbf Q^{(U)})$
$\forall \sigma\in\group{S}_d$.  Finally, it is easy to prove that
under the condition Eq.~\eqref{eq:obscompatsym}, $R_{i}$ satisfy
Eq.  (\ref{eq:obsperminvproperty}). \qed
\end{Proof}

The advantage of using the relabeling symmetry is the reduction of the
number of independent parameters of the generalized quantum
instrument.
Combining Eq. (\ref{eq:diagpart}) with Eq. (\ref{eq:obsperminvproperty}) we have that

\begin{align}
  R'_{i, \vec j} = R'_{\sigma(i), \sigma(\vec j)}.
  \label{eq:rijrsisj}
\end{align}

 Let us now define the equivalence relation between strings
$i, \vec j$ and $i',\vec j'$ as
\begin{equation}
  i, \vec j\sim i', \vec j'\quad\Leftrightarrow\quad i=\sigma(i')\wedge \vec j=\sigma(\vec j'),
\end{equation}
for some permutation $\sigma$. Thanks to Eq.
(\ref{eq:rijrsisj}) there are only as many independent
$R'_{i,\vec{j}}$ as there are equivalence classes among sequences
$i,\vec{j}$. In the simplest case of $N=1$ and arbitrary dimension
$d\geq 2$, there are only two classes, which we denote by $xx$ and
$xy$. The reason is that for any couple $i,j$ there is either a
permutation $\sigma$ such that $\sigma(i),\sigma(j)=1,1$ or
$\sigma(i),\sigma(j)=1,2$, thus the classes are defined by the
conditions $i=j$ or $i\neq j$, respectively. For the case $N=2$ the
vector $i,\vec j$ has three components.  Then there are four or five
equivalence classes depending on the dimension $d$ being $d=2$ or
$d>2$, respectively. We denote these equivalence classes by $xxx, xxy,
xyx, xyy, xyz$ and the set of these elements by $\Rel{3}$.  In the
general case, it is clear that 
the number of classes is given by the number of disjoint partitions of
a set with cardinality $N+1$, with number $p$ of parts $p\leq d$
\cite{note}.


It is useful to introduce the notation
\begin{equation} \label{eq:defrxy}
 R_{x,\vec y}:=R'_{i,\vec j}=R'_{\sigma(i),\sigma(\vec j)},
\end{equation}
where $(x,\vec y)$ is a string of indices that
represents one equivalence class. We will denote by $\defset L$ the
set of equivalence classes $\defset L:=\{(x,\vec y)\}$
and we will use letters from the beginning of the alphabet to name arbitrary element in $\defset L$ in situations, when $N$ is fixed. For example when $N=1$ $(a,b)\in \defset L\equiv\{(x,x),(x,y)\}$.

As a consequence of lemma \ref{lem:covnet} the Eq.~\eqref{eq:obscommutr} can be written as
  \begin{align}
    [R_{x,\vec y}, {U^*}_{\defset{out}} \otimes U^{\otimes
      N}_{\defset{in}}]=0.
  \end{align}
 By Schur's lemmas this implies
the following structure for the operators $R_{x,\vec y}$
\begin{equation}\label{eq:irrdecomp}
  R_{x,\vec y}=\bigoplus_\nu P^\nu \otimes r^\nu_{x,\vec y},
\end{equation}
where $\nu$ labels the irreducible representations in the
Clebsch-Gordan series of ${U^*}_{\defset{out}}\otimes {U}^{\otimes
  N}_{\defset{in}}$, and $P^\nu$ acts as the identity on the invariant
subspaces $\hilb H_\nu$ of the representations $\nu$, while
$r^\nu_{x,\vec y}$ acts on the multiplicity space $\mathbb C^{m_\nu}$
of the same representation.

In the simplest case $N=1$ we have
\begin{equation}
  R_{a,b}=P^p r_{a,b}^p  + P^q  r_{a,b}^q,
  \label{eq:rxxxy}
\end{equation}
where
\begin{equation}
  P^p:=\frac1d|\omega\>\<\omega|, \quad P^q:=(I-P^p)
\end{equation}
and $r_{a,b}^p$ and $r_{a,b}^q$ are non-negative numbers due to  $R_{a,b}\geq 0$. In the case
$N=2$ we have two different decompositions,
depending on whether $d=2$ or $d>2$. In the former case,  we have 
\begin{align}\label{eq:obsdecompor21dim2}
  R_{x,\vec y} = P^\alpha\otimes r^\alpha_{x,\vec y} +
  P^\beta r_{x,\vec y}^\beta,
\end{align}
where $r_{x,\vec y}^\alpha$ is a positive $2 \times 2$ matrix,
while $r_{x,\vec y}^\beta$ is a non-negative real number. The
projections $P^\xi$ on the invariant spaces of the representation
$U^*\otimes U\otimes U$ are the following
\begin{align}
  &P^\alpha\otimes\ketbra ij=\sum_{m=1}^d|\Psi^i_m\> \< \Psi^j_m|,\quad i,j\in\{+,-\}\nonumber\\
  &P^\beta=I\otimes P^+-P^\alpha\otimes\ketbra++,
\end{align}
where $\ket{\Psi^\pm_m}=(|\omega\kk \Ket m\pm \Ket m|\omega\kk)/[2(d\pm1)]^\frac12$,
and $P^+$, $P^-$, are the projections onto the symmetric and
antisymmetric subspace, respectively. When $d>2$, on the other hand, we have
\begin{align}\label{eq:obsdecompor21}
  R_{x,\vec y} = P^\alpha \otimes r_{x,\vec y}^\alpha+
  P^\beta r_{x,\vec y}^\beta + P^\gamma r_{x,\vec y}^\gamma,
\end{align}
where $r_{x,\vec y}^\alpha$ is a positive 2$\times$2 matrix,
while $r_{x,\vec y}^\beta$ and $r_{x,\vec y}^\gamma$ are
non-negative real numbers. The projections $P^\xi$ on the invariant
spaces of the representation $U^*\otimes U\otimes U$ are the following
\begin{align}
  &P^\alpha\otimes\ketbra ab=\sum_{m=1}^d|\Psi^a_m\> \< \Psi^b_m|,\quad a,b\in\{+,-\}\nonumber\\
  &P^\beta=I\otimes P^+-P^\alpha\otimes\ketbra++,\nonumber\\
  &P^\gamma=I\otimes P^--P^\alpha\otimes\ketbra--.
\end{align}
The introduced symmetries have a deep influence on the structure of the replicated POVM as we show in the following lemma.

\begin{lemma}
  \label{lem:povmstruc}
  The properties (\ref{eq:obsdiagcomb}), (\ref{eq:obscommutr}) and
  (\ref{eq:obsperminvproperty}) induce the following structure of the
  replicated POVM's:
  \begin{align} \label{eq:diagonalformforthePOVM}
        G^{(U)}_{i} =
\lambda  U \ketbra{i}{i} U^\dagger +
\frac{1 - \lambda }{d} I,
  \end{align}
which can be seen as a random mixture of a perfect replica with a
trivial measurement (i.e. a measurement producing equiprobably any of
the outcomes) with mixing coefficient $\lambda$,
which is a function of $\bf{R}$.
\end{lemma}
\begin{Proof}
Because of the property (\ref{eq:covpovmform2})
it is sufficient to prove the statement for $U = I $.
Since
$(G^{(I)}_{i})^T = \sum_{\vec{j}} \bra{\vec{j}} R_{i,\vec{j}} \ket{\vec{j}}$
 (see Eq. (\ref{eq:reppovmdiagr})) we have:
\begin{align}
& \bra{k} G^{(I)}_{i} \ket{l}
=\bra{l}\sum_{\vec{j}} \bra{\vec{j}}
R_{i,\vec{j}} \ket{\vec{j}}\ket{k}=
\label{diagonalpovm}
\\
& =
\Tr \left[ \sum_{\vec{j}}R_{i,\vec{j}}
\int \!\! \d U
 \, U^{\otimes N} \otimes U^*
\ket{\vec{j}k}
\bra{\vec{j}l}
(U^{\otimes N} \otimes U^*)^\dagger
\right] \nonumber \\
&=
\Tr \left[\sum_{\vec{j}} R ^{\theta_{2N}}_{i,\vec{j}}
\left(
\int \!\! \d U
 \, U^{\otimes N+1 }
\ket{\vec{j}l}
\bra{\vec{j}k}
U^{\dagger \otimes N+1}
\right)^{\theta_{2N}}
\right],
\nonumber
\end{align}
where we used the property (\ref{eq:obscommutr}) in the equality
\eqref{diagonalpovm} and
$\theta_{2N}$
denotes the partial transpose on $\hilb{H}_{2N}$.
Thanks to the Schur's lemmas we have
\begin{align}
\int \!\! \d U
 \, U^{\otimes N+1 }
\ket{\vec{j}l}
\bra{\vec{j}k}
U^{\dagger \otimes N+1}
=
\sum_{\nu}P^{\nu} \otimes O^{\nu}_{\vec{j},l,k},
\nonumber
\end{align}
where
\begin{align}
 O^{\nu}_{\vec{j},l,k} =
\Tr_{\hilb{H}_\nu} \left[
(P^{\nu} \otimes I^{m_\nu})
\ket{\vec{j} l}
\bra{\vec{j} k}
\right]. \nonumber
\end{align}
We now notice that for $k\neq l$ $\{\vec{j},k \}$
and $\{ \vec{j},l \}$ are two different sets of indices and then there
exists no permutation $\mathsf{S}$ such that
 $\bra{\vec{j},k} \mathsf{S} \ket{\vec{j},l} \neq 0$.
Since any operator of the form
$P^\nu \otimes A $ , $A \in \mathbb{C}^{m_\nu}$ can be written as a
linear combination of permutations
$P^\nu \otimes A = \sum_n a_n \mathsf{S}_n$ we have
\begin{align} \label{eq:offdiagnull}
  \Tr \left[
(P^{\nu} \otimes A)
\ket{\vec{j} l}
\bra{\vec{j} k}
\right] =
\bra{\vec{j} k} \sum_n a_n \mathsf{S}_n \ket{\vec{j} l} = 0
\end{align}
for $k \neq l$.
From Eq. \eqref{eq:offdiagnull} it follows for $\forall k \neq l$ that
$ O^{\nu}_{\vec{j},l,k} = 0$ and hence also
\begin{align}\label{eq:diagformpovm1}
\bra{k} G^{(I)}_{i} \ket{l} = 0  \quad \forall k \neq l
\Rightarrow   G^{(I)}_{i} = \sum_n g^{i}_n \ketbra{n}{n}    
\end{align}
Reminding Eq. (\ref{eq:obspermseedproperty}) we have
\begin{align*}
G^{(I)}_{i} =
 T_\sigma G^{(I)}_{i} T^\dagger_\sigma
\qquad \forall \sigma\in\group{S}_d \mbox{  s.t. }
\sigma(i) = i.
\end{align*}
This implies
\begin{align}\label{eq:permutPOVM}
  \bra{k} G^{(I)}_{i} \ket{k} =
\bra{l} G^{(I)}_{i} \ket{l} \quad
\forall k,l \neq i.
 \end{align}
Eq. \eqref{eq:permutPOVM} combined with Eq.
\eqref{eq:diagformpovm1} and (\ref{eq:obspermseedproperty}) finally leads to
\begin{align}\label{eq:lambda2}
          G^{(I)}_{i} =
\lambda   \ketbra{i}{i}  +
\frac{1 - \lambda }{d} I \qquad 0 \leq \lambda \leq 1.
\end{align}
where $\lambda$ is a function of $\bf{R}$.
Rewriting Eq. (\ref{eq:lambda2}) one has
\begin{align}
\label{eq:lambda}
\lambda =(d\bra{i} G^{(I)}_{i} \ket{i}-1)/(d-1).
\end{align}

Let us note that $\bra{i} G^{(I)}_{i} \ket{i}$ has the same value independently of $i$.
\qed
\end{Proof}

We have shown that the optimization can be restricted without lost of
generality to learning networks obeying Eqs. (\ref{eq:obsdiagcomb}),
(\ref{eq:obscommutr}) and (\ref{eq:obsperminvproperty}). Further in
the paper we always assume that all the considered networks have the
aforementioned properties. This allows us to express the figure of
merit $D[\boldsymbol{\mathcal R}]$ in a different form that will be
more useful for calculations. The expression
\eqref{eq:diagonalformforthePOVM} for the replicated POVM allows us to
write
\begin{align*}
 & D[\boldsymbol{\mathcal R}]=\int \d U \mathscr{D}(  \mathbf E^{(U)} , \mathbf G^{(U)}  ) = \\
&=
(1-\lambda)^2
 \sum_i \int \d U \d \psi \left | \bra{\psi}
 \left( U \ketbra{i}{i} U^\dagger - \frac{1}{d} I \right)
 \ket{\psi} \right |^2 = \\
&=(1-\lambda)^2\sum_i \int \d U
 |\bra{0} U \ket{i}|^4
-
\frac{2}{d}|\bra{0} U \ket{i}|^2
+
\frac{1}{d^2} =  \\
&
= \frac{d-1}{d(d+1)}(1-\lambda)^2
\end{align*}
It is now clear that minimization of the figure of merit
$D[\boldsymbol{\mathcal R}]$ is equivalent to the maximization of
parameter $\lambda = \lambda[\boldsymbol{\mathcal R}]$, which is by Eq.  (\ref{eq:lambda}) directly
related to the maximization of the following quantity:
\begin{equation}\label{eq:deff}
  F[\boldsymbol{\mathcal R}]:=\frac{1}{d}\sum_{i=1}^d \bra{i} G^{(I)}_{i} \ket{i}\equiv \bra{j} G^{(I)}_{j} \ket{j} \quad \forall j
\end{equation}
The relation of $D[\boldsymbol{\mathcal R}]$ and
$F[\boldsymbol{\mathcal R}]$ is given by the following equation
\begin{equation}\label{eq:reldf}
D[\boldsymbol{\mathcal R}]=\frac{d}{d^2-1}(1-F[\boldsymbol{\mathcal R}])^2 .
\end{equation}

The quantity $F[\boldsymbol{\mathcal R}]$, which we actually need to
maximize can be finally written using Eqs.
(\ref{eq:deff}),(\ref{eq:defrxy}),(\ref{eq:reppovmdiagr}) as
\begin{eqnarray}
  F[\boldsymbol{\mathcal R}]&=&\frac{1}{d}\sum_i\sum_{\vec{j}}  \bra{i}_{\defset{out}}\bra{\vec{j}}_{\defset{in}}
  R'_{i,\vec j}\ket{\vec{j}}_{\defset{in}}\ket{i}_{\defset{out}} \nonumber\\
  &=&\frac1{d}\sum_{(x,\vec y)\in\defset L} n(x,\vec y)\<R_{x,\vec y}\>,
  \label{eq:finfigm}
\end{eqnarray}
where $n(x,\vec y)$ is the cardinality of the equivalence class
denoted by the couple $(x,\vec y)$, and $\<R_{x,\vec y}\>=\bra
{i}\bra{\vec j}R'_{i,\vec j}\ket{i}\ket{\vec j}$ for any string
$i,\vec j$ in the equivalence class denoted by $(x,\vec y)$.

\section{Optimization}\label{learn}

In this section we derive optimal quantum learning of a von
Neumann measurement 
 for the scenarios analyzed in the following subsections.

\subsection{$1 \to 1$ Learning}
Suppose that today we are provided with a single use of
a measurement device, and we need its replica to measure a state that
will be prepared only tomorrow. 
Such a scenario is described by the following
scheme.
\begin{align}\label{eq:obs11learn}
  \begin{aligned}
    \Qcircuit @C=0.7em @R=1em {
      &&&\ustick{2}& \ghost{\;\;\;}\\
      \multiprepareC{1}{\;\;\;} &\ustick{0} \qw & \measureD{\mathbf{E}^{(U)}} & \ustick{1} \cw &\pureghost{\;\;\;} \cw\\
      \pureghost{\;\;\;}& \qw& \qw&\qw & \multigate{-2}{\;\;\;}\\
    }
  \end{aligned}
\end{align}
Using the labeling from Eq. (\ref{eq:obs11learn}) and 
the results of Section \ref{sec:obssymm-repl-netw} for $N=1$, 
we have
\begin{align}
  &\defset L=\{(x,x),(x,y)\},\nonumber\\
  &R_{{i}_{210}}=\ket i\bra i_{1}\otimes R_{{x,x}_{20}}+(I-\ket i\bra i)_1\otimes R_{{x,y}_{20}} \label{eq:obsrifromrij}\nonumber\\
  & R_{a,b}=P^p r_{a,b}^p + P^q r_{a,b}^q,\quad (a,b)\in \defset L
\end{align}
We use the identity $\bra i\bra j P^p\ket i\ket j=\delta_{ij}1/d$, $n(x,x)=d$ and $n(x,y)=d(d-1)$, to rewrite the figure of
merit in Eq.~\eqref{eq:finfigm} as
\begin{eqnarray}
  F&=& \<R_{x,x}\>+(d-1)\<R_{x,y}\> \nonumber\\
  &=&\sum_{\nu \in \{p, q \}}\left(
  r_{x,x}^{\nu} \Delta_{x,x}^{\nu} + (d-1) r_{x,y}^{\nu} \Delta_{x,y}^{\nu}\right),
\label{eq:obsfigMeritL11}
\end{eqnarray}
where $\Delta_{x,x}^{p}=\frac1d$, $\Delta_{x,y}^{p}=0$, and
$\Delta^q_{a,b}=1-\Delta^p_{a,b}$.  Let us now write the normalization
conditions for the generalized instrument in terms of operators
$R'_{i,j}$.  We have that that $R_\Omega := \sum_i R_i$ has to be the
Choi-Jamio\l kowski operator of a deterministic quantum network and
must satisfy Eq.  (\ref{eq:recnorm2}), that is
\begin{align}\label{eq:obsnorm11lear}
  R_\Omega = I_2 \otimes I_1 \otimes \rho \qquad \Tr[\rho]=1, \quad
  \rho\geq 0.
\end{align}
The commutation relation (\ref{eq:obscommutr}) implies $[\rho,U] = 0$
and consequently the Schur's lemma requires $\rho =\frac1d I$.
 We take this into account in Eq. (\ref{eq:obsnorm11lear}) and with the help of Eq. (\ref{eq:obsrifromrij}) we get
\begin{equation}
  I_1\otimes R_{x,x}+(d-1)I_1\otimes R_{x,y}=\frac Id,
  \label{eq:obsnormL11}
\end{equation}
which can be equivalently written as (see Eq.(\ref{eq:obsrifromrij}))
\begin{align}
  &r_{x,x}^p+(d-1)r^p_{x,y} = r_{x,x}^q+(d-1)r^q_{x,y} = \frac1d.
  \label{eq:obsnormcond11final}
\end{align}
The above constraint implies the following bound
\begin{align}\label{eq:obsmaxfidelity}
  F =& \sum_{\nu}\left(r_{x,x}^{\nu} \Delta_{x,x}^{\nu} +
  (d-1) r_{x,y}^{\nu} \Delta_{x,y}^{\nu}\right) \le\nonumber\\
  &\sum_{\nu \in \{p, q \}} \overline{\Delta}^{\nu} \left(
    r_{x,x}^{\nu} + (d-1) r_{x,y}^{\nu} \right) = \frac{d+1}{d^2},
\end{align}
where $\overline{\Delta}^{\nu} := \max_{(a,b)\in \defset L} \Delta_{a,b}^{\nu}$.
This bound is achieved by
\begin{align}
  r_{x,x}^q=r_{x,y}^{p}=0, \quad
  r_{x,x}^p=\frac{1}{d},\quad
  r_{x,y}^{q}=\frac{1}{d(d-1)},\nonumber
\end{align}
which corresponds to a generalized instrument
\begin{align}
  R_i&=\ketbra ii_1\otimes\frac1d P^p +(I-\ketbra ii)_{1}\otimes \frac1{d(d-1)}P^{q},
\end{align}
that replicates the original Von Neuman measurement as 
\begin{align}
  &G^{(U)}_i=(R^{i}*E^{(U)}_{10})^T=\nonumber\\
  &\frac{1}{d(d-1)} U\ketbra{i}{i}_{1}U^{\dagger} +
  \frac{d^2-d-1}{d^2(d-1)}I.
\end{align}
Based on Eq. (\ref{eq:reldf}) we conclude that the optimal value of
$D[\boldsymbol{\mathcal R}]$ achieved by the aforementioned network is
\begin{equation}
D_{opt}=\frac{d}{d^2 -1}(1-\frac{d+1}{d^2})^2 .
\end{equation}

The optimal learning strategy can be realized by the following network
\begin{align}
  \begin{aligned}
    \Qcircuit @C=1em @R=1em {
      &                                   &  &            \ustick{2}&
      \ghost{\mathbf{P}}         &&\\
      &\multiprepareC{1}{\frac{1}{d}\ketbra{\omega}{\omega}}&\ustick{A_1}\qw&\qw& \multimeasureD{-1}{\mathbf{P}}
      &   \multipuregate{1}{f} \cw\\
      &\pureghost{\frac{1}{d}\ketbra{\omega}{\omega}}& \ustick{0} \qw &
      \measureD{\mathbf{E}^{(U)}}  & \ustick{1}  \cw&  \pureghost{f} \cw
       }
  \end{aligned}
\end{align}
that operates as follows. The storing part of the strategy consists of
preparing maximally entangled state
$\frac{1}{d}\ketbra{\omega}{\omega}$ and measuring one part of it by
the unknown measurement that we want to learn. Application of the
learned POVM on some system $\hilb{H}_2$ is achieved by measuring
two outcome POVM
 $\mathbf{P}:=\{P^p,P^q\}$ on the system $\hilb{H}_2$ and on the
 unmeasured part of the state
$\frac{1}{d}\ketbra{\omega}{\omega}$.
The last step of the optimal learming strategy
consists in a classical processing $f$
of the outcome $k$ of $\mathbf{E}^{(U)}$
and of the outcome $n$ of $\mathbf{P}$.
The function $f$
that produces the actual outcome of the replicated measurement is
defined as follows
\begin{align}
  \label{eq:2}
  f(k,n) =
\left \{
  \begin{array}{lcr}
    k & \mbox{if} & n=p \\
    j \neq k & \mbox{if} & n=q
  \end{array}
\right.
\end{align}
where the outcome $j$ in the second case is randomly generated with
flat distribution.

When the outcome $n=p$ of the measurement $\vec P$ occurs, we achieved a teleportation, of input state of $\hilb{H}_2$ to the past, that is to the system $\hilb{H}_2$. In this sense the optimal $1\mapsto 1$ Learning is achieved using the probabilistic teleportation [ref!!!].
We stress that the optimal scheme differs from the one in which one
optimally estimates $\mathbf{E}^{(U)}$ and then reproduces the
estimated POVM. 
In contrast to the optimal learning of unitaries, it is possible to prove that the 
\emph{optimal estimate \& prepare} strategy for measurements achieves strictly lower
performance than the strategy derived in this section.

\subsection{$2 \to 1$ Learning} 

We now consider the case in which we have two uses of the unknown Von Neumann measurement at our  
disposal
\begin{align}\label{eq:obs21learn}
  \begin{aligned}
    \Qcircuit @C=0.7em @R=1em { &&&&&&& \ustick{4}& \ghost{\;\;\;}\\
      \multiprepareC{1}{\;\;\;} &\ustick{0} \qw & \measureD{\mathbf{E}^{(U)}} &
      \ustick{1} \cw & \pureghost{\;\;\;} \cw & \ustick{2} \qw &
      \measureD{\mathbf{E}^{(U)}} & \ustick{3} \cw &
      \pureghost{\;\;\;} \cw \\
      \pureghost{\;\;\;}& \qw& \qw&\qw & \multigate{-1}{\;\;\;}&\qw &
      \qw&\qw &
      \multigate{-2}{\;\;\;} \\
    }
  \end{aligned}
\end{align}
As a consequence of the symmetries
introduced in Section \ref{sec:obssymm-repl-netw} we have
\begin{align}
  &\defset L=\{(x,xx),(x,xy),(x,yx),(x,yy),(x,yz)\}\nonumber\\
  &R_i = \sum_{j,k} \ketbra{j}{j}_3 \otimes \ketbra{k}{k}_1 \otimes R'_{i,jk}  \label{eq:obslear21strucr}\\
  &[R_{i,jk}, U^*_4 \otimes U_2 \otimes U_0] = 0 \label{eq:obsler21comr}\\
  &R'_{i,jk} = \left\{\begin{array}{lcl}
      R_{x,xx} & {\rm if}& i=j=k \\
      R_{x,xy} & {\rm if}& i=j\neq k \\
      R_{x,yx} & {\rm if}& i=k\neq j \\
      R_{x,yy} & {\rm if}& j=k\neq i \\
      R_{x,yz} & {\rm if}& i\neq j \neq k \neq i.
    \end{array}
  \right.
  \label{eq:obslear21perminv}
\end{align}
The figure of merit (\ref{eq:finfigm}) becomes
\begin{align}
  F = \frac1d\sum_{(a,bc)\in\defset L}n(a,bc)\<R_{a,bc}\>.
\end{align}
Let us now consider the normalization condition of the  optimal 
generalized instrument 
\begin{align}
  \sum_iR_i = I_4 \otimes I_3 \otimes S_{210} \qquad \Tr_{2}[S]=
  I_1 \otimes \rho_0.
\end{align}
Thank to Eq. (\ref{eq:obslear21strucr}) we have
\begin{align}
  & \sum_iR_i = \sum_{i, j,k} \ketbra{j}{j}_3 \otimes \ketbra{k}{k}_1
  \otimes R'_{i,jk} =
  I_4 \otimes I_3 \otimes S_{210} \nonumber \\
  &\sum_{i, k} \ketbra{k}{k}_1 \otimes R'_{i,jk} = I_4 \otimes
  S_{210},
  \quad \forall j \nonumber \\
  &\sum_{i} R'_{i,jk} = I_4 \otimes \bra{k}S_{210}\ket{k}_1,\quad
  \forall j,k
\end{align}
Using the property (\ref{eq:rijrsisj}) we obtain
\begin{align}
  &I_4  \otimes \bra{k}S_{210}\ket{k}_1 =   \sum_{i} R'_{i,jk} = \sum_{i} R'_{\sigma(i),\sigma(j)\sigma(k)} = \nonumber\\
  &=I_4  \otimes (\bra{k} T^\dagger_\sigma) S_{210} (T_\sigma\ket{k}_1)\quad \forall j,k \; ,
\end{align}
which implies   
\begin{align}
  \sum_{i} R'_{i,jk} = I_4 \otimes T_{20} \;\; \forall j,k \qquad
  \Tr_{20}[T] = 1. \label{eq:obsnorminner}
\end{align}
The commutation relation  (\ref{eq:obscommutr}) implies
$  [I_4 \otimes  T_{20}, U^*_4 \otimes U_2 \otimes U_0] = 0$
and by taking the trace on  $\hilb{H}_4$ we get
\begin{align}\label{eq:obscommutt}
  [T_{20}, U_0 \otimes U_2] = 0,
\end{align}
which due to Schur's Lemmas requires $T_{20} = t_+P^+ +t_-P^- $.
The normalization $\Tr_{20}[T]=1$ becomes
\begin{align}
  \label{eq:tplus}
d_+t_++d_-t_- =1,
\end{align}
where $d_\pm\equiv \Tr[P^\pm]$
 and Eq. (\ref{eq:obsnorminner}) now reads for all $j,k$
\begin{align}
  &\sum_{i} R'_{i,jk} =I_4 \otimes (t_+P^+ + t_-P^-) =\nonumber\\
  &t_+(P^{\alpha}\otimes  \ketbra ++ + P^\beta) + t_-( P^{\alpha} \otimes \ketbra -- + P^\gamma).
  \label{eq:normpar}
\end{align}

As a consequence of Eq. (\ref{eq:obsnorminner}) the optimal strategy can be parallelized.
\begin{align}\label{eq:obs21learnpar}
  \begin{aligned}
    \Qcircuit @C=0.7em @R=1em {
      &&&\ustick{4}&\ghost{\;\;\;}\\
      \multiprepareC{2}{\;\;\;}&\ustick{0} \qw&\measureD{\mathbf{E}^{(U)}} &\ustick{1}\cw& \pureghost{\;\;\;}\cw\\
      \pureghost{\;\;\;}&\ustick{2}\qw&\measureD{\mathbf{E}^{(U)}} &\ustick{3}\cw& \pureghost{\;\;\;}\cw\\
      \pureghost{\;\;\;}&\qw&\qw&\qw&\multigate{-3}{\;\;\;} }
  \end{aligned}
\end{align}
Eq. (\ref{eq:obs21learnpar}) provides a further symmetry of the %
problem:
\begin{lemma}\label{lem:obs21perminv1}
  The operator $R'_{i,jk}$ in Eq. (\ref{eq:obslear21strucr}) can be
  chosen to satisfy:
  \begin{align}\label{eq:obspermsymmetry}
    R'_{i,jk} = \mathsf{S}R'_{i,kj}\mathsf{S} \quad \forall k,j
  \end{align}
  where $\mathsf{S}$ is the swap operator
  $\mathsf{S}\ket{k}_2\ket{j}_0 = \ket{j}_2\ket{k}_0$.
\end{lemma}
\begin{Proof}
  The proof consists in the standard averaging argument.  let us
  define $\overline{R}_{i,jk}:=\frac12 (R'_{i,jk} + \mathsf{S}
  R'_{i,kj} \mathsf{S})$.  It is easy to prove that $\{
  \overline{R}_{i,jk} \}$ satisfies the normalization
  (\ref{eq:obsnorminner}) and that gives the same value of
  $F[\boldsymbol{\mathcal R}]$ as $R'_{i,kj}$.\qed
\end{Proof}

Eq. (\ref{eq:obspermsymmetry}) together with the decomposition
(\ref{eq:obsdecompor21}) gives for $\forall (a,bc)\in \defset L$
\begin{align}\label{eq:obssimplifiedr}
  &\sigma_z r_{a,bc}^{\alpha} \sigma_z = r_{a,cb}^{\alpha} \quad
  r_{a,bc}^{\beta} = r_{a,cb}^{\beta} \quad
  r_{a,bc}^{\gamma}=r_{a,cb}^{\gamma}
\end{align}
where $\sigma_z = \begin{pmatrix}1 & 0 \\0& -1\end{pmatrix}$.

Considering that $n(x,xx)=d$, $n(x,xy)=n(x,yx)=n(x,yy)=d(d-1)$, and
$n(x,yz)=d(d-1)(d-2)$, and that $\mathsf SR_{x,xy}\mathsf S=R_{x,yx}$, the figure of
merit in Eq.~\eqref{eq:finfigm} can be written as
\begin{align}\label{eq:obsfigmer21redux}
  F =&\<R_{x,xx}\>+(d-1)\<R_{x,yy}\> + 2(d-1)\<R_{x,xy}\> + \nonumber\\
  &(d-1)(d-2) \<R_{x,yz}\> =\nonumber \\
  & = \sum_{\nu} \Tr[\Delta_{x,xx}^\nu r_{x,xx}^\nu +(d-1)\Delta_{x,yy}^\nu r_{x,yy}^\nu +\nonumber\\
  & 2(d-1) \Delta_{x,xy}^\nu r_{x,xy}^\nu +(d-1)(d-2)
  \Delta_{x,yz}^\nu r_{x,yz}^\nu]
\end{align}
where
\begin{equation}
  \Delta_{a,bc}^\nu :=\Tr_{\hilb{H}_\nu}[\ketbra{ijk}{ijk}],
  \label{eq:deltas}
\end{equation}
and $i,jk$ is any triple of indices in the class denoted by $a,bc$.
Notice that in the case $d=2$ the last term in the sum of
Eq.~\eqref{eq:obsfigmer21redux} is 0.

\begin{figure}[h]
    \includegraphics[width=8cm ]{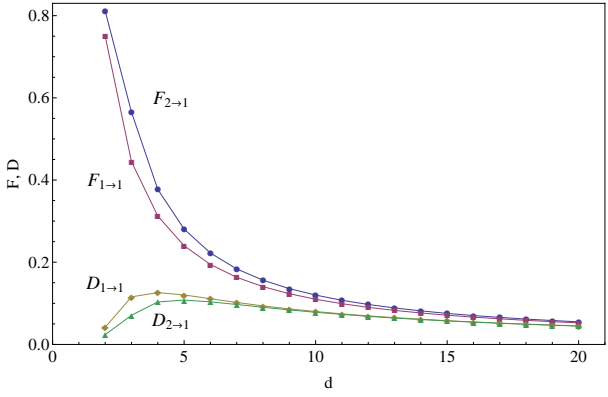}
    \caption{\label{fig:learnobs} Optimal learning of a measurement
      device: we present the values of $D,F$ for different values of
      the dimension $d$.  The squared dots represent the optimal
      learning from a single use ($1 \to 1$ learning) while the round
      dots and triangles represent the optimal learning from two uses
      ($2 \to 1$ learning).  }
\end{figure}

The optimization of $F[\boldsymbol{\mathcal R}]$
can be carried out in two steps:
first we maximize $F[\boldsymbol{\mathcal R}]$
for any fixed value of $t_+$ that satisfies
Eq. \eqref{eq:tplus}; finally we optimize the value of
$t_+$.
The optimization of $F[\boldsymbol{\mathcal R}]$
for fixed $t_+$
is carried out in Appendix \ref{app21}. According to Eq.~\eqref{eq:fide}
we can write the figure of merit as
\begin{align}\label{eq:obs21figmerasfunctionoft}
  F[\boldsymbol{\mathcal R}] &= \frac{d^2 + 3d}{2(d+1)}t_+ +
  \frac{\sqrt{(d-1)t_+t_-}}{\sqrt{d+1}} + \frac{d}2 t_-
\end{align}
The last step of the optimization can be easily done by making the
substitution $t_- = d_-^{-1}(1-d_+t_+)$ in Eq.
(\ref{eq:obs21figmerasfunctionoft}) and then maximizing $F = F(t_+)$.
We will omit the details of the derivation and we rather show a plot
(Fig.~\ref{fig:learnobs}) representing the values of $D,F$ depending
on the dimension.

\begin{figure}[h]
  \includegraphics[width=8cm ]{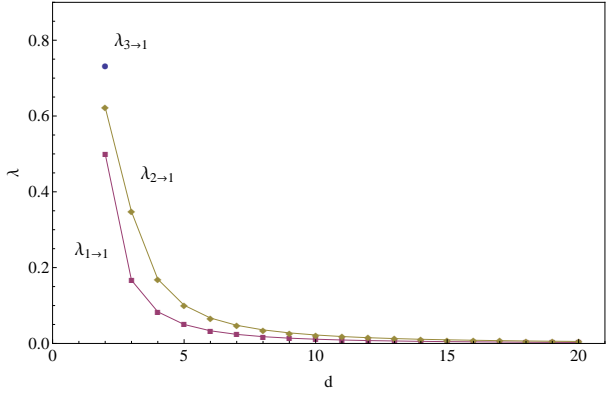}
  \caption{\label{fig:learnobs2} Optimal learning of a measurement
    device: we present the values of $\lambda$, the admixture of
    perfect replica to white noise in the produced measurement for
    different values of the dimension $d$.  The squared dots represent
    the optimal learning from a single use ($1 \to 1$ learning) while
    the diamonds represent the optimal learning from two uses ($2 \to
    1$ learning).  }
\end{figure}
Due to Lemma \ref{lem:povmstruc} the replicated POVM has the following
form:
\begin{eqnarray}
  G_i^{(U)}&=&\lambda E_i^{(U)} +(1-\lambda) \frac{1}{d}I \nonumber \\
  &=& \frac{dF-1}{d-1}U \ketbra{i}{i}U^\dagger +\frac{1-F}{d-1}I, \nonumber
\end{eqnarray}
where the values of the coefficient $\lambda$ describing the random mixing of a perfect replica with a trivial measurement are depicted on Figure \ref{fig:learnobs2}.

\subsection{$3 \to 1$ learning}
In this section we consider a learning network, which
exploits $3$ uses of the measurement device
and  produces a  single replica:
\begin{align}\label{eq:obslera31generalschem}
  \begin{aligned}
    \Qcircuit @C=0.5em @R=1em {
      &&&&&&&&&&& \ustick{6}& \ghost{\;\;\;}\\
      \multiprepareC{1}{\;\;\;} &\ustick{0} \qw & \measureD{\mathbf{E}^{(U)}} &
      \ustick{1} \cw &\pureghost{\;\;\;} \cw & \ustick{2} \qw &
      \measureD{\mathbf{E}^{(U)}} & \ustick{3} \cw &\pureghost{\;\;\;} \cw &
      \ustick{4} \qw &
      \measureD{\mathbf{E}^{(U)}} & \ustick{5} \cw &\pureghost{\;\;\;} \cw  \\
      \pureghost{\;\;\;}& \qw & \qw&\qw &\multigate{-1}{\;\;\;}&\qw &
      \qw&\qw &\multigate{-1}{\;\;\;}&\qw & \qw&\qw &
      \multigate{-2}{\;\;\;}}
  \end{aligned}\ .
\end{align}
In order to simplify the problem we restrict ourselves to the qubit
case, that is we set $d=2$. The derivation of the optimal learning
network turns out to be very involved although it follows the same
lines as for the $2 \to 1$ case. We made the calculations analytically with the help of a symbolic mathematical program. 

The $3 \to 1$
scenario deserves interest because the optimal solution 
does not allow a strategy having the $3$ uses of the measurement device in
parallel. In other words the optimal strategy needs to be adaptive.

Let us consider the normalization condition for the generalized
instrument $\{ R_i \}$:
\begin{align}
  &\sum_{ijkl}\ketbra{jkl}{jkl}_{531} \otimes R_{i,jkl} = I_{65}
  \otimes S_{43210} \nonumber\\
  &\Tr_{4}[S]= I_3 \otimes T_{210}
  \label{eq:norm3to1}
\end{align}
This implies
\begin{align}
  &\sum_{i} R_{i,jkl} = I_{6}\otimes
  \bra{kl}S_{43210}\ket{kl}_{31}\quad \forall j,\nonumber\\
  &\bra{kl}\Tr_4[S]\ket{kl}= \bra{l}T\ket{l}_1\quad\forall k.
\end{align}
From the relabeling symmetry $R_{i,jkl} =
R_{\sigma(i),\sigma(j)\sigma(k)\sigma(l)}$ we have $\bra{kl}S\ket{kl}
= \bra{\sigma(k)\sigma(l)}S\ket{\sigma(k)\sigma(l)}$, and consequently
\begin{align}
  \bra{kl}\Tr_4[S]\ket{kl}_{31} = \frac1{d^2}\Tr_{431}[S]=:\widetilde{T}_{20},\quad\forall k,l.
\end{align}
This fact along with Eq.~\eqref{eq:norm3to1} allows us to conclude
that
\begin{align}
  \Tr_4[S] =& \Tr_4\left[\sum_{kl}\ketbra{kl}{kl}_{31}\otimes \bra{kl}S_{43210}\ket{kl} \right] = \nonumber \\
  &\sum_{kl}\ketbra{kl}{kl}_{31}\otimes \widetilde{T}_{20} = I_{31} \otimes \widetilde{T}_{20}
\end{align}
 which means that the first two uses can be in parallel.  We
notice that in general $\bra{kl}S\ket{kl} =
\bra{\sigma(k)\sigma(l)}S\ket{\sigma(k)\sigma(l)}$ does not imply that
$\bra{ kl}S\ket{kl} = \widetilde{S}$ is independent of $k,l$, but only
that $\bra{ kl}S\ket{kl} = \widetilde{S}_{ab}$, where $a,b$ denotes
the equivalence class of the couple $(k,l)$. Consequently, we cannot
in general assume that all the examples can be used in parallel. In
fact, the optimal learning network has the following causal structure
\begin{align}\label{eq:obs31learnseq}
  \begin{aligned}
    \Qcircuit @C=0.7em @R=1em { \multiprepareC{2}{\;\;\;}&\ustick{0}
      \qw&\measureD{\mathbf{E}^{(U)}} &\ustick{1}\cw& \pureghost{\;\;\;} \cw&
      &&\ustick{6}&
      \ghost{\;\;\;}\\
      \pureghost{\;\;\;}&\ustick{2}\qw&\measureD{\mathbf{E}^{(U)}}
      &\ustick{3}\cw& \pureghost{\;\;\;} \cw& \ustick{4}
      \qw&\measureD{\mathbf{E}^{(U)}} &\ustick{5}\cw&
      \pureghost{\;\;\;}\cw\\
      \pureghost{\;\;\;}&\qw&\qw&\qw& \multigate{-2}{\;\;\;}& \qw&\qw
      &\qw& \multigate{-2}{\;\;\;} }
  \end{aligned}\ .
\end{align}
where the state of system $4$ depends on the classical outcome in
system $3$ and $1$. The optimal value of $F[\boldsymbol{\mathcal R}]$
is approximately $0,87$ (we remind that for the $1 \to 1$ learning we
had $F = 0,75$, while for the $2 \to 1$ case we had $F= 0,81$). The
corresponding value of coefficient $\lambda$ (see
Eqs.(\ref{eq:lambda}),(\ref{eq:deff})) are depicted on Fig.
(\ref{fig:learnobs2}).
\begin{remark}
  One can wonder whether without assuming any symmetry it is possible
  to find a non-symmetric parallel strategy $\{R_i\}$ that achieves
  the optimal value of $F[\boldsymbol{\mathcal R}]$.  However we
  remind that for any strategy $\{R_i\}$ we can build a symmetric one
  with the same normalization, that is without spoiling the
  parallelism, and giving the same fidelity. Since the optimal
  symmetric network cannot be parallel, we have that any other optimal
  network has to be sequential as well.
\end{remark}


\section{Conclusions}\label{conc}

We analyzed optimal learning of a measurement device. Our approach to
the problem is based on the formalism of quantum combs and generalized
quantum instruments, introduced in Refs.
\cite{comblong,memoryeff,architecture}.
The original problem can be significantly simplified by utilizing
the symmetries provided by the figure of merit. In particular,
covariance and relabeling symmetry allow us to 
significantly decrease the number of parameters, without affecting the
figure of merit.
As a consequence of the symmetry of the learning network the
replicated measurement can be seen as a random mixture of a perfect
replica of the measurement device to be learnt with weight $\lambda$
and of a trivial measurement producing all possible outcomes with the
same probability independently of the input state with weight
$1-\lambda$.
For $2 \to 1$ and $3 \to 1$ learning the first two uses of the unknown
measurement device can be parallelized, and
and 
this result can be generalized to $N \to 1$ learning. However, the
optimal learning algorithm cannot be further parallelized, namely the
examples exceeding the second one must be used sequentially. This
feature is very unusual, and it occurs in few cases of quantum
algorithms \cite{Watrousdiscrimini,fiuramicu}.
For example, while the quantum part of Shor's algorithm
can be parallelized, Grover's algorithm cannot, as was proved in Ref.
\cite{zalka}. Our results prove that quantum learning of a von Neumann
measurement shares with Grover's algorithm the impossibility of
parallelizing without affecting optimality. The parallelization of the
first two examples from this point of view is a curious exception.

An obvious extension of the work would be to study the scaling of the performance of the optimal learning strategy with respect to $N$. However,
 our results show that optimal learning networks with different $N$ do not share the same the initial steps. This means that the optimization of $N \rightarrow 1$ learning can not be done inductively building on the results from $N-1 \rightarrow 1$ case. The complexity of the optimization in general case rises mainly due to the causal influence of steps of the learning strategy on the remaining part of the network, which is reflected in the recursive structure of the normalization constraints.


\section*{Acknowledgments}
This work has been supported by the European Union through FP$7$ STREP project COQUIT and by the Italian Ministry of Education through grant PRIN 2008
Quantum Circuit Architecture.

\appendix

\section{Calculations for $2 \to 1$ Learning}
\label{app21}

The explicit expression of $\Delta^\nu_{a,bc}$ in
Eq.~\eqref{eq:deltas} is given by
\begin{align} \label{eq:obsdeltas} &\Delta_{x,xx}^\alpha=
  \begin{pmatrix}
    \frac{2}{d+1}& 0 \nonumber\\
    0 & 0
  \end{pmatrix}, \quad\Delta_{x,xy}^\alpha= \frac12
  \begin{pmatrix}
    \frac{1}{d+1} & \frac{1}{\sqrt{d^2-1}} \nonumber\\
    \frac{1}{\sqrt{d^2-1}} & \frac{1}{d-1}
  \end{pmatrix},\nonumber\\
  &\Delta_{x,yy}^\alpha= \Delta_{x,yz}^\alpha= 0,\quad \Delta_{x,yx}^\alpha=\sigma_z \Delta_{x,xy}^\alpha \sigma_z  \nonumber\\
  &\Delta_{x,xx}^\beta = \frac{d-1}{d+1},\quad \Delta_{x,xy}^\beta =
  \Delta_{x,yx}^\beta =
  \frac{d}{2(d+1)},\nonumber\\
  & \Delta_{x,yy}^\beta = 1,\quad \Delta_{x,yz}^\beta = \frac12, \nonumber\\
  &\Delta_{x,xx}^\gamma = \Delta_{x,yy}^\gamma = 0, \quad
  \Delta_{x,xy}^\gamma = \Delta_{x,xy}^\gamma =\frac{d-2}{2(d-1)},\nonumber \\
  &\Delta_{x,yz}^\gamma = \frac12.
\end{align}
Introducing the notation
\begin{align}\label{eq:obs21defnizdirridot}
  &s_{x,xx}^{\nu} := r_{x,xx}^{\nu}&&s_{x,xy}^{\nu} :=  (d-1)r_{x,xy}^{\nu} \nonumber\\
  &s_{x,yx}^{\nu} := (d-1)r_{x,yx}^{\nu} & &s_{x,yy}^{\nu} :=
  (d-1)r_{x,yy}^{\nu}\nonumber\\
  & s_{x,yz}^{\nu} := (d-2)(d-1)r_{x,yz}^{\nu},
\end{align}
the figure of merit (\ref{eq:obsfigmer21redux}) becomes
\begin{align} %
  F &= F_\alpha + F_\beta + F_\gamma \nonumber\\
  & F_\nu \equiv \sum_{(a,bc)\in \defset L}\Tr[\Delta_{a,bc}^\nu
  s_{a,bc}^\nu], \quad \nu\in\{\alpha,\beta\gamma\}
\end{align}
We express $R'_{i,jk}$ through $R_{a,bc}$ $(a,bc)\in \defset L$ and
Equation (\ref{eq:obsdecompor21dim2}). Depending on $j=k$ or $j\neq k$
Eq.~\eqref{eq:normpar} is equivalent to the following relations
\begin{align}\label{eq:obsnorm21ultrafinal}
&j=k \Rightarrow \nonumber \\
  & s_{x,xx}^{\alpha} + s_{x,yy}^{\alpha} =
  \begin{pmatrix}
    t_+ & 0 \\
    0 & t_-
  \end{pmatrix}
  \nonumber\\
  &s_{x,xx}^{\beta} +   s_{x,yy}^{\beta} = t_+\nonumber\\
  &s_{x,xx}^{\gamma} +   s_{x,yy}^{\gamma} = t_-  \nonumber \\
&j\neq k \Rightarrow \nonumber \\
  &s_{x,xy}^{\alpha}+ \sigma_z s_{x,xy}^{\alpha}
  \sigma_z + s_{x,yz}^{\alpha} =
  \begin{pmatrix}
    (d-1)t_+ & 0 \\
    0 & (d-1)t_-
  \end{pmatrix}
  \nonumber \\
  &2s_{x,xy}^{\beta}+ s_{x,yz}^{\beta} = (d-1)t_+\\
  &2s_{x,xy}^{\gamma}+s_{x,yz}^{\gamma} = (d-1)t_- ,
\end{align}
where we utilized Equation (\ref{eq:obssimplifiedr}) implied by Lemma \ref{lem:obs21perminv1}.
We now derive the optimal learning network 
for a fixed value of $t_+$ (remember that $t_- = (1-d_+t_+ )/d_-$).

First we maximize $F_\beta$ and $F_\gamma$
for the case $d \geq 3$. Using the expressions for the $\Delta_{i,jk}^\nu$ from Eq. (\ref{eq:obsdeltas}) we have:
\begin{align}\label{eq:obsboundfbeta}
  F_\beta =& \sum_{(a,bc)\in \defset L}\Tr[ \Delta_{a,bc}^\beta s_{a,bc}^\beta]\leq
  \max(\Delta_{x,xx}^\beta, \Delta_{x,yy}^\beta)t_+ + \nonumber\\
& + \max(\Delta_{x,xy}^\beta, \Delta_{x,yz}^\beta)(d-1)t_+ =\nonumber \\
  &= \Delta_{x,yy}^\beta t_+ + \Delta_{x,yz}^\beta(d-1)t_+= \nonumber \\
  & = t_+ + \frac{(d-1)t_+}{2}= \frac{(d+1)t_+}{2}
\end{align}
and
\begin{align}\label{eq:obsboundfgamma}
  F_\gamma =& \sum_{(a,bc)\in \defset L}\Tr[ \Delta_{a,bc}^\gamma s_{a,bc}^\gamma]\leq
  \max(\Delta_{x,xx}^\gamma, \Delta_{x,yy}^\gamma)t_- +\nonumber\\
 &+ \max(\Delta_{x,xy}^\gamma, \Delta_{x,yz}^\gamma)(d-1)t_- =\nonumber\\
  & = \Delta_{x,yz}^\gamma (d-1)t_- = \frac{(d-1)t_-}{2}.
\end{align}
where we used the normalizations constraints~(\ref{eq:obsnorm21ultrafinal}).
The upper bounds (\ref{eq:obsboundfbeta}) and ~(\ref{eq:obsboundfgamma})
can be achieved by taking
\begin{align}
&s_{x,xx}^{\beta}=s_{x,xy}^{\beta}=s_{x,yx}^{\beta}=s_{x,xx}^{\gamma}=s_{x,xy}^{\gamma}=s_{x,yx}^{\gamma}=0, \nonumber \\
&s_{x,yy}^{\beta}=t_+, \quad s_{x,yz}^{\beta}=(d-1)t_+, \nonumber \\
&s_{x,yy}^{\gamma}=t_-, \quad s_{x,yz}^{\gamma}=(d-1)t_-. \nonumber
\end{align}
For $d=2$ the irreducible representation denoted by $\gamma$
and the $x,yz$ class do not exist and the optimization
yields $s_{x,xy}^\beta = t_+ (d-1)$.

Let us now consider $F_\alpha$
(in this case there is no difference between $d\geq 3$ and $d=2$). 
Based on the expression of $\Delta_{i,jk}^\alpha$ we have:
\begin{align}
  F_\alpha =& \sum_{(a,bc)\in \defset L}\Tr[\Delta_{a,bc}^\alpha s_{a,bc}^\alpha]= \nonumber \\
  & \Tr[\Delta_{x,xx}^\alpha s_{x,xx}^\alpha] + \Tr[
  \Delta_{x,xy}^\alpha s_{x,xy}^\alpha] +\Tr[
  \Delta_{x,yx}^\alpha s_{x,yx}^\alpha]  =
  \nonumber \\
  & \Tr\left[ \begin{pmatrix}
      \frac2{d+1} & 0 \\
      0 & 0
  \end{pmatrix}
s_{x,xx}^\alpha +
\begin{pmatrix}
    \frac{1}{d+1} &\frac{1}{\sqrt{d^2-1}}  \\
    \frac{1}{\sqrt{d^2-1}} &  \frac{1}{d-1}
  \end{pmatrix}
 s_{x,xy}^\alpha \right] \leq \nonumber \\
&
    \frac2{d+1}
t_+
 +
\Tr \left[
  \begin{pmatrix}
    \frac{1}{d+1} &\frac{1}{\sqrt{d^2-1}}  \\
    \frac{1}{\sqrt{d^2-1}} &  \frac{1}{d-1}
  \end{pmatrix}
 s_{x,xy}^\alpha \right],
\label{eq:obsboundfalfa}
\end{align}
and the bound can be achieved by taking
\begin{align}
  s_{x,xx}^\alpha =
  \begin{pmatrix}
    t_+ & 0 \\
    0 & t_-
  \end{pmatrix}.
\end{align}

Let us now focus on the expression $\Tr[  \Delta_{x,xy}^\alpha s_{x,xy}^\alpha]$.
The normalization constraint (\ref{eq:obsnorm21ultrafinal})
  for the operator $s_{x,xy}^\alpha$ can be rewritten as:
\begin{align}\label{eq:obs21normforalpha}
  &  s_{x,yz}^{\alpha,+,-} = s_{x,yz}^{\alpha,-,+} = 0 \nonumber \\
  & s_{x,yz}^{\alpha,+,+}+  2s_{x,xy}^{\alpha,+,+} =    (d-1)t_+ \nonumber \\
  & s_{x,yz}^{\alpha,-,-}+ 2s_{x,xy}^{\alpha,-,-} = (d-1)t_-,
\end{align}
where we denoted $s_{a,bc}^{\alpha,\pm,\pm}:=\bra{\pm}s_{a,bc}^{\alpha}\ket{\pm}$. 
Then we have
\begin{align}
  \Tr[ \Delta_{x,xy}^\alpha s_{x,xy}^\alpha] &=
  \frac{s_{x,xy}^{\alpha,+,+}}{d+1} +
  \frac{s_{x,xy}^{\alpha,+,-}}{\sqrt{d^2-1}}+\nonumber\\
  &\frac{s_{x,xy}^{\alpha,-,+}}{\sqrt{d^2-1}}+
  \frac{s_{x,xy}^{\alpha,-,-}}{d-1} \leq  \nonumber \\
  &\frac{ s_{x,xy}^{\alpha,+,+}}{d+1}+
  2\frac{\sqrt{s_{x,xy}^{\alpha,+,+}
      s_{x,xy}^{\alpha,-,-}}}{\sqrt{d^2-1}} +
  \frac{s_{x,xy}^{\alpha,-,-}}{d-1} \leq   \label{eq:obs21usolapos} \\
  &\frac{(d-1)t_+}{2(d+1)} +
  \frac{\sqrt{(d-1)t_+t_-}}{\sqrt{d+1}}+ \frac{t_-}{2}
  \label{eq:obs21usoilconst}
\end{align}
where we used the positivity of the operator $s_{x,xy}^\alpha$ for the
inequality (\ref{eq:obs21usolapos}) and the normalization
(\ref{eq:obs21normforalpha}) for the second inequality
(\ref{eq:obs21usoilconst}). The upper bound in Eq.
(\ref{eq:obs21usoilconst}) can be achieved by taking
\begin{align}
  s_{x,xy}^\alpha =
  \frac{(d-1)}2   \begin{pmatrix}
    t_+ & \sqrt{t_+t_-}\\
    \sqrt{t_+t_-} & t_-
  \end{pmatrix}
\end{align}
Finally, combining the optimal values of $F_\alpha$, $F_\beta$, and $F_\gamma$ we have
\begin{equation}\label{eq:fide}
F[\boldsymbol{\mathcal R}]=
\frac{d^2 + 3d}{2(d+1)}t_+ +
  \frac{\sqrt{(d-1)t_+t_-}}{\sqrt{d+1}} + \frac{d}2 t_-
\end{equation}


\begin{thebibliography}{99}
\bibitem{directsum}
This is  equivalent to a usage of a direct sum over the classical outcomes.
\bibitem{note} For $N+1\geq d$, this number is known as Bell number
$B_{N+1}$.
In the case $N+1<d$ the solution is provided by the sum for
$k=1,\dots,d$ of numbers of disjoint partitions of a set with $N+1$
elements into $k$ subsets, which is the sum of Stirling numbers of the
second kind $S(N+1,k)$ for $1\leq k\leq d$. 
\end{thebibliography}
\end{document}